\documentstyle{paper}
\begin{document}

\title{GRAVITATIONAL WAVES FROM INSPIRALLING COMPACT BINARIES: A
POST-NEWTONIAN APPROACH}

\author{ Clifford M. WILL \\
{\it  McDonnell Center for the Space Sciences, Department of Physics,\\
Washington University, St. Louis, Missouri 63130 USA} \\
Preprint No. WUGRAV-94-3 }

\maketitle

\section*{Abstract}

Inspiralling binary systems of neutron stars or black holes are
promising sources of gravitational radiation detectable by large-scale
laser interferometric gravitational observatories, such as the US LIGO
and Italian-French VIRGO projects.  Accurate theoretical gravitational-waveform
templates will be needed to carry out matched filtering
data analysis of the detectors' output once they are on the air by the
end of this decade.  For all but the final, strongly general
relativistic coalescence of the two bodies, high-order post-Newtonian
methods are playing a major role in the theorists' efforts to develop
the needed templates.  This paper discusses the foundations of this
method, and provides a compendium of useful formulae and results.

\section{Introduction}

Binary systems containing compact objects such as neutron stars or
black holes, which undergo a rapid inspiral toward a final state of
coalescence because of gravitational-radiation damping have become the
focus of intense theoretical scrutiny.  This is because they are now
viewed as the most promising sources of gravitational waves for
kilometer-class laser-interferometric gravitational observatories
such as the US LIGO and the Italian-French
VIRGO projects (Abramovici, {\it et al.} 1992).
The gravitational waves emitted by binary neutron
stars
in the last few minutes of
inspiral can have characteristic wave amplitudes of $h \approx
10^{-22}$ out to 200 Mpc, with frequencies sweeping through the
detectors' sensitive bandwidth from 10 Hz to 1000 Hz.  For details on
LIGO/VIRGO plans, see the article by Thorne in this volume.

Strategies are being studied for extracting the characteristic
``chirp'' signal from the noisy data in an array of detectors, and
using the signal to obtain useful information about the sources, such
as their direction, distance, masses, spins, and so on.  These
strategies usually involve matched filtering of a set of theoretical
``template waveforms'' against the output of the detectors (Thorne
1987, Schutz 1991, Krolak {\it et al.} 1991,
Jotania {\it et al.} 1992, Finn 1992).

As early as 1963, Dyson (1963) considered the problem of two coalescing
neutron stars, estimating the energy emitted in gravitational waves,
and the characteristic frequency.  During the 1970's the problem was
reconsidered in more detail, and estimates of gravitational-wave
and neutrino emission, tidal disruption, and mass ejection were made
(Clark and Eardley 1977, Lattimer and Schramm 1974, 1976).
More recently, the possibility of measuring cosmological
parameters, such as the Hubble parameter, using data from coalescing
binaries has been considered (Schutz 1986, Krolak and Schutz 1987,
Chernoff and Finn 1993,
Markovi\'c 1993).  The rate of formation of such coalescing
systems has been studied by
several authors (Clark, {\it et al.} 1979, Narayan, {\it et al.} 1991,
Phinney 1991).  The most recent estimates take into account
the existence of binary pulsar systems in our
galaxy, such as the
Hulse-Taylor system PSR 1913+16, and the system PSR 1534+12, both with
gravitational-radiation damping lifetimes short compared to the age of
the galaxy, and arrive at a rate of the order of three coalescences
per year out to 200 Mpc.  Many of the key issues in coalescing
binaries and their detection by LIGO/VIRGO systems have been
summarized recently by the Caltech group (Cutler {\it et al.} 1993).

The importance of these systems poses a strong challenge to theorists:
calculate with high accuracy and reliability the orbital evolution
and gravitational-wave emission from inspiralling binaries, beginning
from the time when the orbital frequency reaches about 10 Hz, through
to coalescence.  Depending on the system in question, a variety of
theoretical tools must be brought to bear on the problem, including
post-Newtonian approximations, numerical relativity, and black-hole
and neutron-star
perturbation theory.

To
illustrate the importance and relevance of various theoretical
approaches, we consider three ``benchmark'' systems: two 1.4 $M_\odot$
neutron stars, two 10 $M_\odot$ black holes,
and a system containing one of each (Table 1).
It is common to assume that gravitational
radiation reaction acting during the past history of the binary system
has circularized the orbits; this will certainly be the case, for
example, for the binary pulsar PSR 1913+16.
(High-eccentricity ``capture'' orbits, which remain eccentric
throughout the inspiral have also been discussed (Lincoln and Will
1990).)

\vspace{1pc}

Table 1.  Characteristics of Inspiralling Binaries
\vspace{.5pc}

\begin{tabular}{lccc} \\ \hline
Type of Binary& $2\times 1.4M_\odot$ & $10M_\odot+
1.4M_\odot$ & $2\times 10M_\odot$ \\ \hline
$r/m$ when $f$=10Hz&180&70 &47   \\
$r/m$ when $f$=1000Hz&8&3&2  \\
$r/m$ at plunge& 6.0 & 5.5 & 6.0      \\
$f$ at plunge (Hz)     & 1,400 & 360 & 190     \\
Time to coalescence (s)& 1,030 & 270 & 40      \\
Number of GW cycles to coalescence& 16,500 & 3,520 & 600      \\   \hline
\end{tabular}

\vspace{1pc}

When the gravitational-wave signal enters the sensitive bandwidth of
the detectors, at around $f \approx 10$  Hz, the system is undergoing a
quasi-circular, adiabatic inspiral.  Since $m/r \approx v^2$ is small,
where $m$ is the total mass of
the system, and $r$ and $v$ are the orbital separation and velocity,
post-Newtonian
methods can be used to describe the orbit and gravitational waveform
(we use units in which $G=c=1$).
However, because $m/r$ and $v^2$
ultimately grow to values that may not be so small,
higher-order post-Newtonian
effects must be taken into account.  This is especially true in the
evolution of the orbital frequency under the influence of radiation
reaction, because the theoretical template must
match the observed signal to within a fraction of a phase over the
hundreds to thousands of gravitational wave cycles observed in the
sensitive bandwidth (Cutler {\it et al.} 1993).

At around $6.0\, m$ for the equal-mass systems, and $5.5\,m$ for the 1.4:10
system, the motion changes from
an adiabatic inspiral to an unstable plunge (Kidder {\it et al.} 1992,
1993a) (we work in harmonic
coordinates, in which the event horizon of an isolated nonrotating black hole
of
mass $m$ is
at radius $m$ instead of $2m$, and the innermost stable circular
orbit for a test body is at $5m$).
Because the bodies are compact, tidal
effects can generally be ignored until the final plunge (Bildsten and
Cutler 1992).  After the plunge,
numerical relativity must come into play, because post-Newtonian
methods fail, yet the system is too highly distorted and dynamical for
perturbation techniques to be reliable.  For systems containing neutron
stars, Newtonian or general relativistic hydrodynamics computer
codes must be employed (for recent work and references to earlier work
see, for example Rasio and Shapiro 1992, Shibata {\it et al.} 1992,
Centrella and McMillan 1993); for
two black holes, vacuum codes with horizon boundary conditions must be
used (see, for example Detweiler and Blackburn 1992,
Anninos {\it et al.} 1993).

At some stage, the final coalesced
object, be it a black hole or a neutron star, will settle down to a
state that can be approximated as a perturbation of a
stationary system, and will oscillate
in its characteristic set of quasi-normal modes.  Many of the details
of these modes have been well studied by perturbation theory and
numerical general relativity (see Andersson 1992 for recent results
and references to earlier work on black-hole quasi-normal modes).

For the three benchmark systems, Table 1 lists approximate values of
relevant quantities, such as the orbital separation when the system
enters the sensitive bandwidth, when the gravitational wave
frequency reaches 1 kHz, and when the transition to plunge occurs,
and the time elapsed and the number of cycles of
gravitational waves emitted from detection to coalescence.
It is important to note that, because of the increase in
laser shot noise above 100 Hz in LIGO/VIRGO detectors, a typical
double neutron star system at 200 Mpc will leave the sensitive bandwidth at
a few hundred Hz, well
before plunge and coalescence.  It may be possible to study the
high-frequency waves
from the final
coalescence of such systems using special-purpose
narrow-band techniques, such as ``dual recycling''
(Krolak {\it et al.} 1993).  For the stronger yet lower-frequency massive
black-hole systems, it
should be possible to observe the entire coalescence, from inspiral to
quasi-normal mode radiation using broad-band detection.

This paper will focus on post-Newtonian methods that have been used to
study the inspiral phase and the emitted signal.  A post-Newtonian
approximation is an expansion of corrections to Newtonian gravitational
theory in terms of a small parameter $\epsilon \approx m/r \approx
v^2$.  The three key ingredients are the equations of motion, the
gravitational waveform, and formulae for energy and angular-momentum
flux.
Schematically the equations of motion are given as
\begin{equation}
d^2 {\bf x}/dt^2 = -(m{\bf x}/r^3)
[1+O(\epsilon)+O({\epsilon}^{3/2})+O({\epsilon}^2)
 +O({\epsilon}^{5/2})+ \dots ]\,,
\end{equation}
where $\bf x$ and $r=\vert \bf x \vert$ denote the separation vector
and distance
between the bodies, and $m=m_1 + m_2$ denotes the
total mass.
The quantity relevant for detectors is the gravitational waveform,
the transverse-traceless (TT) part of the far-zone field, denoted
$h^{ij}$.  In terms of an expansion
beyond the quadrupole formula, it has the schematic form,
\begin{equation}
h^{ij} = {2 \over R} \left\{ Q^{ij} [ 1 + O(\epsilon^{1/2}) + O(\epsilon)
+ O(\epsilon^{3/2}) + \dots ] \right\} _{TT} \,,
\end{equation}
where $Q^{ij}$ represents the usual quadrupole term (two time
derivatives of
the mass quadrupole moment tensor).  Finally, the flux of energy and
angular momentum, which can be calculated from $h^{ij}$, are
intimately related to the gravitational radiation reaction terms in
the equations of motion, and are important for calculating the
secular evolution of the orbit during the inspiral.  Schematically,
they can be written,
\begin{equation}
dE/dt=(dE/dt)_Q [1 +O(\epsilon) + O(\epsilon^{3/2}) + \dots ] \,,
\end{equation}
\begin{equation}
d{\bf J}/dt=(d{\bf J}/dt)_Q [1 +O(\epsilon) + O(\epsilon^{3/2}) + \dots ]
\,.
\end{equation}
Tables 2, 3 and 4 show the significance of the various terms in these
expansions, and provide references to recent work on the higher-order
terms.

\vfill
\break

Table 2. Post-Newtonian Equations of Motion

\begin{tabular}{lll} \\ \hline
Order &Effect&Reference\\ \hline
$\epsilon$&Post-Newtonian&Standard texts\\
$\epsilon^{3/2}$&Spin-orbit&Standard texts\\
$\epsilon^2$&Post$^2$-Newtonian&Standard texts\\
&Spin-spin&Standard texts\\
$\epsilon^{5/2}$&Radiation damping&Standard texts\\
$\epsilon^3$&Post$^3$-Newtonian& ?\\
$\epsilon^{7/2}$&PN damping&Iyer \& Will (1993)\\
$\epsilon^{4}$&Spin-orbit damping&Iyer \& Will (1994)\\ \hline
\end{tabular}

\vspace{1pc}

Table 3.  Post-Newtonian Gravitational Waveforms

\begin{tabular}{lll} \\ \hline
Order &Effect&Reference\\ \hline
$\epsilon^{1/2} \& \epsilon$&Post-Newtonian&Wagoner \& Will (1976)\\
$\epsilon^{3/2}$&Post-Newtonian&Wiseman (1992)\\
&Tails&Blanchet \& Damour (1992), Wiseman (1993)\\
&Spin-orbit&Kidder, {\it et al.} (1993b), Kidder (1994) \\
$\epsilon^2$&Post$^2$-Newtonian&Blanchet, {\it et al.} (1994)\\
&PN Tails&Blanchet, {\it et al.} (1994)\\
&Spin-spin&Kidder, {\it et al.} (1993b), Kidder (1994)\\
$\epsilon^{5/2}$&Non-linear memory&Christodoulou (1991), Wiseman \& Will
(1991),\\
&& Thorne (1992)\\ \hline
\end{tabular}

\vspace{1pc}

Table 4.  Post-Newtonian Energy and Angular Momentum Flux

\begin{tabular}{lll} \\ \hline
Order &Effect&Reference\\ \hline
$\epsilon$&Post-Newtonian&$E$: Wagoner \& Will(1976), Blanchet \&
Sch\"afer (1989)\\
&&${\bf J}$: Junker \& Sch\"afer (1992)\\
$\epsilon^{3/2}$&Tails&$E$: Blanchet \& Damour (1992), Wiseman (1993)\\
&Spin-orbit&$E,\, {\bf J}$: Kidder, {\it et al.} (1993b), Kidder
(1994)\\
$\epsilon^2$&Post$^2$-Newtonian&Blanchet, {\it et al.} (1994)\\
&PN Tails&Blanchet, {\it et al.} (1994)\\
&Spin-spin&$E,\, {\bf J}$: Kidder, {\it et al.} (1993b), Kidder
(1994)\\ \hline
\end{tabular}

\vspace{.5pc}

\section{Equations of Motion}

Equations of motion for two bodies of arbitrary mass and spin have been
developed by many authors (for reviews and references see
Damour (1987), Lincoln and Will (1990)).
By eliminating the center of mass of the system, it is possible to convert the
two-body
equations of motion to a relative one-body equation of motion given by
\begin{equation}
{\bf a} = {\bf a}_N + {\bf a}_{PN}^{(1)} + {\bf a}_{SO}^{(3/2)} +
{\bf a}_{2PN}^{(2)}
+ {\bf a}_{SS}^{(2)} + {\bf a}_{RR}^{(5/2)}
 + O({\bf a}^{(3)}) \,,
\end{equation}
where the subscripts denote the nature of the term, post-Newtonian
(PN), spin-orbit (SO), post-post-Newtonian (2PN), spin-spin (SS), and
radiation reaction (RR); and
the superscripts denote the order in $\epsilon$.  The individual terms
are given by
\begin{eqnarray}
{\bf a}_N = && - {m \over r^2} {\bf \hat n} \,, \\
{\bf a}_{PN}^{(1)} = && - {m \over r^2} \biggl\{  {\bf \hat n} \left[
-2(2+\eta)
{m \over r} + (1+3\eta)v^2 - {3 \over 2} \eta \dot r^2 \right]
  -2(2-\eta) \dot r {\bf v} \biggr\} \,, \\
{\bf a}_{SO}^{(3/2)} = && {1 \over r^3} \biggl\{ 6 {\bf \hat n} [( {\bf \hat
n} \times
{\bf v} ) {\bf \cdot} ( 2{\bf S} + {{\delta m} \over m}{\bf \Delta} )] -
[ {\bf v} \times
(7
{\bf S} + 3 {{\delta m} \over m} {\bf \Delta} )] \nonumber \\
&& + 3 \dot r [ {\bf \hat n} \times (3 {\bf S}
+ {{\delta m} \over m} {\bf \Delta} )] \biggr\} \,, \\
{\bf a}_{2PN}^{(2)} = && - {m \over r^2} \biggl\{ {\bf \hat n} \biggl[ {3
\over 4}
(12+29\eta) ( {m \over r} )^2 + \eta(3-4\eta)v^4 + {15 \over 8}
\eta(1-3\eta)
\dot r^4 \nonumber \\
 && - {3 \over 2} \eta(3-4\eta)v^2 \dot r^2
- {1 \over 2} \eta(13-4\eta) {m \over r} v^2 - (2+25\eta+2\eta^2)
{m \over
r} \dot r^2 \biggr] \nonumber \\
 && - {1 \over 2} \dot r {\bf v} \left[ \eta(15+4\eta)v^2 -
(4+41\eta+8\eta^2)
{m \over r} -3\eta(3+2\eta) \dot r^2 \right] \biggr\} \,, \\
{\bf a}_{SS}^{(2)} = && - {3 \over \mu r^4} \biggl\{ {\bf \hat n} ({\bf S_1
\cdot
S_2}) + {\bf S_1} ({\bf \hat n \cdot S_2}) + {\bf S_2} ({\bf \hat n
\cdot
S_1}) - 5 {\bf \hat n} ({\bf \hat n \cdot S_1})({\bf \hat n \cdot
S_2})
\biggr\}  \,, \\
{\bf a}_{RR}^{(5/2)}= && {8 \over 5} \eta {m^2 \over r^3} \left\{ \dot r
{\bf \hat n}
\left[ 18v^2 + {2 \over 3} {m \over r}-25 \dot r^2  \right] - {\bf v}
\left[ 6v^2 - 2 {m \over r}-15 \dot r^2 \right] \right\} \,,
\end{eqnarray}
where ${\bf x \equiv x_1-x_2}$, ${\bf v}=d{\bf x}/dt$, ${\bf \hat n}
\equiv {\bf x}/r$, $\mu \equiv m_1m_2/m$, $\eta \equiv \mu/m$, $\delta
m \equiv m_1-m_2$, ${\bf S \equiv S_1+S_2}$, and
${\bf \Delta} \equiv m({\bf S_2}/m_2-{\bf S_1}/m_1)$.
It is useful to note that spin-orbit and spin-spin terms are formally
post-Newtonian corrections,
since the spin of each body is of order $mR_A\bar v$ where $R_A$ is the
size of
body A and $\bar v$ is the rotational velocity of the body, so
that
the spin-orbit and spin-spin accelerations are of order $(R_A/r)v\bar v$
and
$(R_A/r)^2\bar v^2$, respectively, {\it i.e.} $O(\epsilon)$
compared to the Newtonian
acceleration.
For
compact
objects, however, $R_A$ is of order $m_A$, while $\bar v$ could be of
order
unity, so that the spin-orbit and spin-spin accelerations are
{\it effectively} of
(post)$^{3/2}$-Newtonian and (post)$^2$-Newtonian order, respectively.
For application to coalescing compact binaries,
we will adopt this convention for treating spin effects (see Tables 2
-- 4).

Explicit formulae for radiation-reaction terms at higher ``post-Newtonian''
order
(i.e. at higher order in $\epsilon$ beyond the lowest-order term ${\bf
a}_{RR}^{(5/2)}$) have been derived by Iyer and Will (1993, 1994), including
the
post-Newtonian term ${\bf a}_{RR:PN}^{(7/2)}$, and the spin-orbit term
${\bf a}_{RR:SO}^{(4)}$.

Note that, while
${\bf a}_N$, ${\bf a}_{PN}$,
${\bf a}_{2PN}$, and ${\bf a}_{RR}$ are all confined to the orbital
plane,
in general ${\bf a}_{SO}$ and ${\bf a}_{SS}$ (as well as
${\bf a}_{RR:SO}^{(4)}$)
are not.  As a result,
the
orbital plane will precess in space (except for specific spin
orientations), resulting in modulations of the observed gravitational
waveform (Cutler {\it et al.} 1993,
Apostolatos, {\it et al.} 1994, Kidder 1994).

The spin vectors satisfy a set of precession equations of motion,
given by
\begin{eqnarray}
{\bf \dot S_1} = && {1 \over r^3} \left\{ ({\bf L}_N \times {\bf S_1})
\left ( 2+{3
\over 2}
{m_2 \over m_1} \right ) - {\bf S_2 \times S_1} + 3({\bf \hat n \cdot S_2})
{\bf \hat n \times S_1} \right\} \,, \\
{\bf \dot S_2} =  && {1 \over r^3} \left\{ ({\bf L}_N \times {\bf S_2})
\left ( 2+{3
\over 2}
{m_1 \over m_2} \right ) - {\bf S_1 \times S_2} + 3({\bf \hat n \cdot S_1})
{\bf \hat n \times S_2} \right\} \,,
\end{eqnarray}
where ${\bf L}_N=\mu{\bf x \times v}$.
The first term in each expression is the precession due to
spin-orbit
coupling, and the second and third terms are due to spin-spin
coupling.  In terms of ${\bf S}$ and ${\bf \Delta}$, the precession
equations can be written in the form
\begin{eqnarray}
{\bf \dot S} = && {1 \over r^3} \biggl\{ {1 \over 2} {\bf L}_N \times (7{\bf S}
+
3 {{\delta m} \over m} {\bf \Delta})
+ 6 \eta \left[ ({\bf \hat n \cdot S})({\bf \hat n \times S}) - \eta
({\bf \hat n \cdot \Delta})({\bf \hat n \times \Delta}) \right] \nonumber
\\
&& + 3 \eta {{\delta m} \over m} \left[
({\bf \hat n \cdot \Delta })({\bf \hat n \times S}) +
({\bf \hat n \cdot S})({\bf \hat n \times \Delta}) \right]
\biggr\}  \,, \\
{\bf \dot \Delta}= && {1 \over r^3} \biggl\{ {1 \over {2\eta}} {\bf L}_N
\times [(3-5\eta){\bf \Delta} +3{{\delta m} \over m} {\bf S}]
 -{\bf S \times \Delta} \nonumber \\
&&  +3 {{\delta m} \over m} [({\bf \hat n \cdot S})({\bf \hat n \times S})
- \eta
({\bf \hat n \cdot \Delta})({\bf \hat n \times \Delta})] \nonumber \\
&& +3[(1-2\eta)({\bf \hat n \cdot S})({\bf \hat n \times \Delta})-2\eta
({\bf \hat n \cdot \Delta})({\bf \hat n \times S})] \biggr\} \,.
\end{eqnarray}
Note that the precessions of the spins are post-Newtonian effects,
since $L_N/r^3 \approx (v/r)(\mu/r) \approx \epsilon (d/dt)$, and
$S_i/r^3 \approx m_iR \bar v /r^3 \approx \epsilon^{3/2} (d/dt)$.

Through (post)$^2$-Newtonian order, the equations of motion, including
spin effects, can be derived from a generalized Lagrangian.  A
generalized Lagrangian is a
function of the relative position, velocity, and spins, and also of
the acceleration.  To this order, that is, neglecting radiation
reaction, the Lagrangian leads to conserved energy and angular
momentum given by
\begin{equation}
E =  E_N +  E_{PN} + E_{SO} + E_{2PN} + E_{SS} \,,
\end{equation}
\begin{equation}
{\bf J} = {\bf S} +{\bf J}_N + {\bf J}_{PN} + {\bf J}_{SO} + {\bf J}_{2PN} \,,
\end{equation}
where
\begin{eqnarray}
E_N = && \mu \left ( {1 \over 2} v^2 - {m \over r} \right ) \,,  \\
E_{PN} = && \mu \left\{
{3 \over 8}
(1-3\eta) v^4
+{1 \over 2} (3+\eta) v^2 {m \over r} +
{1 \over 2} \eta {m \over r} \dot r^2 + {1 \over 2} ({m \over r})^2
\right\} \,, \\
E_{SO} = && {1 \over r^3} {\bf L}_N \cdot ({\bf S} + {{\delta m}
\over m} {\bf \Delta}) \,, \\
E_{2PN} =  && \mu \biggl\{ {5 \over 16}(1-7\eta+13\eta^2) v^6
+ {1 \over 8} (21-23\eta-27\eta^2) {m \over r} v^4 \nonumber\\
&&+ {1 \over 4} \eta (1-15\eta) {m \over r} v^2 \dot r^2
 - {3 \over 8} \eta (1-3\eta){m \over r} \dot r^4
 - {1 \over 4} (2+15\eta) \left( {m \over r} \right) ^3 \nonumber \\
&&+ {1 \over 8} (14-55\eta+4\eta^2) \left( {m \over r} \right) ^2 v^2
+ {1 \over 8} (4+69\eta+12\eta^2) \left( {m \over r} \right) ^2 \dot r^2
 \biggr\} \,, \\
E_{SS} = && {1 \over r^3} \left\{ 3 \left( {\bf \hat n \cdot S_1}
\right)
  \left( {\bf \hat n \cdot S_2} \right) - \left( {\bf S_1 \cdot S_2}
  \right) \right\} \,,
\end{eqnarray}
\begin{eqnarray}
{\bf J}_N = && {\bf L}_N \,, \\
{\bf J}_{PN} = && {\bf L}_N \left\{
{1 \over 2} v^2
(1-3\eta) + (3+\eta) {m \over r} \right\} \,, \\
{\bf J}_{SO} = && {\mu \over m} \left\{
  {m \over r} {\bf \hat n \times} \left( {\bf \hat n \times}
\left[
  3 {\bf S} +{{\delta m} \over m} {\bf \Delta} \right] \right)  - {1 \over 2}
  {\bf v \times} \left( {\bf v \times} \left[ {\bf S}+ {{\delta m} \over m}
{\bf \Delta}
\right] \right) \right\}
\,, \\
{\bf J}_{PPN} = && {\bf L}_N \biggl\{
{1 \over 2}
(7-10\eta-9\eta^2) {m \over r}
v^2  - {1 \over 2}\eta (2+5\eta) {m \over r} \dot r^2 \nonumber\\ &&
+ {1 \over 4} (14-41\eta+4\eta^2)
\left( {m \over r} \right)^2
+{3 \over 8} (1-7\eta+13\eta^2) v^4 \biggr\} \,.
\end{eqnarray}
Note that there is no spin-spin contribution to {\bf J}.

Kidder {\it et al.} (1992, 1993a) developed a ``hybrid'' set of equations of
motion for spinless bodies,
in which the sum of the terms in Eqs. (6), (7) and (9) that are independent of
$\eta$ is replaced by the exact expression for motion in the
Schwarzschild metric of a body of mass $m$, while the terms dependent
on $\eta$ are left unaffected. The resulting equation of motion is
thus exact in the test-body ($\eta \to 0$) limit, and valid to
(post)$^2$-Newtonian order for arbitrary masses, when suitably
expanded.  Similar hybrid expressions were derived for energy and
angular momentum.  With these equations it was possible to show that,
in the absence of radiation reaction, the innermost stable circular
orbit would vary from $5\,m$ in the test-mass limit, to $6.03\,m$ for
equal masses.

\begin{figure}[t]
   \vspace{15pc}
   \caption{Quasicircular relative orbit for equal masses (left) and
for a 10:1 mass ratio (right).  Marked points denote separation in
units of $m$.  Around $r=6m$ for equal masses and $r=5.5m$ for the
10:1 mass ratio, orbit undergoes an unstable plunge (from Lincoln and
Will 1990).}
\end{figure}

\section{Gravitational Waveform}

The two-body, far-zone gravitational waveform has been derived through
$O(\epsilon^{3/2})$ beyond the quadrupole approximation, i.e. through
(post)$^{3/2}$-Newtonian order, including the effects of spin, and of
the gravitational wave tail.  The latter effect is caused by
scattering of the outgoing gravitational waves off the
Schwarzschild-like curved spacetime around the source, and is a
hereditary effect, that is, it depends on the history of the system
prior to retarded time $u= t-R$.  The most rigorous route to the
waveform is via the Blanchet-Damour-Iyer (BDI) formalism (Blanchet and
Damour 1986, 1988, 1989; Damour and Iyer 1991), in
which the waveform is written in terms of symmetric, trace-free (STF)
multipoles of source densities, in the form
\begin{eqnarray}
h^{ij} =&& {2 \over R} \biggl\{ \stackrel{(2)}{I^{ij}} + {1 \over 3}
\stackrel{(3)\,}{I^{ijk}} N^k + {1 \over 12}
\stackrel{(4) \quad}{I^{ijkl}} N^kN^l
+ {1 \over 60} \stackrel{(5) \qquad}{I^{ijklm}}N^kN^lN^m + \dots \nonumber \\
&& + \epsilon^{kl(i} \biggl[ {4 \over 3} \stackrel{(2)\,}{J^{j)k}}N^l
+ {1 \over 2} \stackrel{(3) \quad}{J ^{j)km}} N^lN^m
+ {2 \over 15}
\stackrel{(4) \qquad}{J^{j)kmn}} N^lN^mN^n + \dots \biggr] \biggr\}_{TT}
\,, \label{waveform}
\end{eqnarray}
where $R$ and $\bf {\hat N}$ are the distance and unit vector from
source to observer.
The notation $(n)$ over each multipole denotes the number of
derivatives with respect to retarded time, $\epsilon^{ijk}$ is the
completely antisymmetric Levi-Civita symbol, and parentheses around
indices denote symmetrization.  The TT part of any tensor $A^{ij}$ is
given by $A_{TT}^{ij}=A^{lm}(P^{il}P^{jm}- {1 \over 2} P^{ij}P^{lm})$,
where $P^{ij}=\delta^{ij}-N^iN^j$.  The $I^{ij\dots}$
are mass multipole moments, while the $J^{ij\dots}$ are current
multipole moments.  Each of the successive
terms in Eq. (\ref{waveform}) is $O(\epsilon^{1/2})$ smaller than
the previous term, with the
current quadrupole moment $J^{ij}$ itself $O(\epsilon^{1/2})$ smaller
than the mass
quadrupole moment $I^{ij}$.  To the order needed for the waveform
accurate to $O(\epsilon^{3/2})$, the moments are given by
\begin{eqnarray}
I^{ij} = && \mu \biggl\{ ( x^i x^j ) \left[ 1 +
{29 \over 42}(1-3\eta)v^2
- {1 \over 7}(5-8\eta){m \over r} \right] - {4 \over 7}(1-3\eta) r
\dot r
(x^iv^j) \nonumber \\ && + {11 \over 21}(1-3\eta) r^2
(v^iv^j) \biggr\}^{STF}
 + {8 \over 3} \eta \left[ x^i
  \left( {\bf v \times \xi} \right)^j \right]^{STF} - {4 \over 3} \eta
 \left[ v^i \left( {\bf x \times \xi} \right)^j \right]^{STF}
\nonumber \\
&& + I_{Tail}^{ij} \,, \\
I_{Tail}^{ij}= && 2m \int_0^\infty \stackrel{(2)}{I^{ij}}(u-u^\prime)
\left [\ln \left ( {u^\prime \over {2s}} \right ) + {11 \over 12} \right ]
du^\prime \,,
\\
I^{ijk} = && - \mu {\delta m \over m} \biggl\{ ( x^i x^j x^k )
\left[ 1
+{1 \over 6} (5 - 19\eta )v^2
- {1 \over 6} (5 - 13\eta ){m \over r} \right] \nonumber \\
&&- (1-2\eta) r \dot r
(x^ix^jv^k) + (1-2\eta) r^2
(x^iv^jv^k) \biggr\}^{STF} \,, \\
I^{ijkl} = && \mu (1-3\eta) \left( x^i x^j x^k x^l \right)^{STF}\,, \\
I^{ijklm} = &&- \mu {\delta m \over m}(1-2\eta)  \left( x^i x^j x^k x^l x^m
\right)^{STF}\,, \\
J^{ij} = && - \mu {\delta m \over m} \biggl\{ x^i ({\bf x
  \times v}) ^j  \left [ 1+ {1 \over 2} (1-5\eta)v^2+2(1+
\eta){m \over r} \right ] \nonumber \\
&& + {1 \over 28} (1-2\eta) \left [ 5r \dot r
v^i- \left ( v^2+2{m \over r} \right ) x^i \right ] ({\bf x \times
v})^j \biggr\}^{STF} - {3 \over 2} \eta ( x^i \Delta^j
  )^{STF}\,, \\
J^{ijk} =  && \mu (1-3\eta) \left[ x^i x^j ({\bf x
  \times v}) ^k \right]^{STF} +  2 \eta (x^i x^j \xi^k)^{STF} \,, \\
J^{ijkl} = && - \mu {\delta m \over m}(1-2\eta)  \left[ x^i x^j x^k ({\bf x
  \times v}) ^l \right]^{STF} \,,
\end{eqnarray}
where ${\bf \xi} \equiv {\bf S}+ (\delta m/m) {\bf \Delta}$.
The parameter $s$ that appears in the ``Tail'' moment is arbitrary;
its origin is the matching between the flat light cones $t-R$=constant
in harmonic coordinates in the near zone, and the logarithmically
corrected light cones $t-R-2m \ln R$=constant in the far zone.  This
matching is to be done at some intermediate, but arbitrary radius $s$,
and leads to a retarded time $u$ in all moments given by
\begin{equation}
u=t-R-2m \ln (R/s)
\end{equation}
To $O(\epsilon^{3/2})$, the $s$ dependence
implicit in $u$ can be shown to offset the explicit
$s$ dependence in the Tail term, so that it can be chosen arbitrarily,
in principle.  Eq. (29) is a purely formal expression, since it is a
divergent integral for general sources; however the relevant quantity
for the waveform,
$(d^2/du^2)I_{Tail}^{ij}$, is convergent.

\begin{figure}[t]
   \vspace{29pc}
   \caption{Gravitational waveform plotted against orbital phase for a
10:1 mass-ratio system, ignoring spin.  Plotted is $(R/2\mu)h_+$, viewed by an
observer in the orbital plane.  Plots begin at an orbital separation of
$15 m$ and terminate around $4 m$ (from Wiseman 1992). }
\end{figure}

The result for the waveform is
\begin{eqnarray}
h^{ij} = && {{2\mu} \over R} \biggl[ \tilde Q^{ij} + {{\delta m} \over m}
P^{1/2}Q^{ij}
+ PQ^{ij} + PQ_{SO}^{ij} + {{\delta m} \over m} P^{3/2}Q^{ij}
 + P^{3/2}Q_{SO}^{ij} \nonumber \\
&& + P^{3/2}Q_{TAIL}^{ij} + O(\epsilon^2)
\biggr]_{TT}
\end{eqnarray}
where, as before, the superscripts denote the effective order of
$\epsilon$, and subscripts label the nature of the term, and where
\begin{eqnarray}
\tilde Q^{ij} = &&  2 \left ( v^iv^j - {m \over r}n^in^j \right )
\,, \\
P^{1/2}Q^{ij} = && 3({\bf \hat N \cdot \hat n})
{m \over
r} \left[ 2n^{(i}v^{j)} - \dot r n^in^j \right] + ({\bf \hat N \cdot
v})
\left[ {m \over r} n^in^j - 2v^iv^j \right] \,, \\
PQ^{ij} = && {1 \over 3} \biggl\{ (1-3\eta) \biggl[
({\bf \hat N \cdot \hat n}
)^2 {m \over r} \left[ (3v^2 - 15\dot r^2 + 7{m \over r})n^in^j +
30\dot r n^{(i}v^{j)} - 14v^iv^j \right] \nonumber \\ &&
+ ({\bf \hat N \cdot \hat n})({\bf \hat N \cdot v}){m \over r} \left[
12\dot r n^in^j - 32n^{(i}v^{j)} \right] + ({\bf \hat N \cdot
v})^2
\left[ 6v^iv^j - 2{m \over r} n^in^j \right] \biggr] \nonumber \\
&& + \left[ 3(1-3\eta)v^2 - 2(2-3\eta){m \over r} \right] v^iv^j +
4 {m \over r} \dot r (5+3\eta)n^{(i}v^{j)} \nonumber \\
&& + {m \over r} \left[ 3(1-3\eta)\dot r^2 - (10+3\eta)v^2 + 29 {m
\over r}
\right] n^in^j \biggr\} \,, \\
PQ_{SO}^{ij} = &&  {2 \over r^2} ({\bf \Delta \times
\hat N})^{(i} n^{j)} \,, \\
P^{3/2}Q^{ij} = && (1-2\eta) \biggl\{ ({\bf \hat N
\cdot \hat n})^3 {m \over r} \biggl[ {5 \over 4}(3v^2-7\dot r^2 +6{m
\over r})
\dot r n^in^j - {17 \over 2} \dot r v^iv^j  \nonumber \\ &&
- {1 \over 6}(21v^2-105\dot r^2 +44{m \over r})n^{(i}v^{j)}
\biggr] \nonumber \\
&& +{1 \over 4} ({\bf
\hat N \cdot \hat n})^2({\bf \hat N \cdot v}) {m \over r} \biggl[
58v^iv^j + (45\dot r^2
-9v^2 -28{m \over r})n^in^j -108 \dot r n^{(i}v^{j)}
 \biggr] \nonumber \\ &&
+ {3 \over 2}({\bf \hat N \cdot \hat n})({\bf \hat N \cdot
v})^2 {m \over r} \biggl[ 10n^{(i}v^{j)}
- 3 \dot r n^in^j \biggr]
+ {1 \over 2} ({\bf \hat N \cdot v})^3 \left[ {m \over r}n^in^j -
4v^iv^j \right] \biggr\}  \nonumber \\ && +
 {1 \over 12} ({\bf \hat N \cdot \hat n}){m
\over r}
\biggl[ 2n^{(i}v^{j)} \left( \dot r^2 (63+54\eta) - {m \over
r}(128-36\eta)
+ v^2(33-18\eta) \right) \nonumber \\ && + n^in^j\dot r \left( \dot
r^2(15
-90\eta)-v^2(63-54\eta)+{m \over r}(242-24\eta) \right)
-\dot r v^iv^j(186+24\eta) \biggr] \nonumber \\ &&
+ ({\bf \hat N \cdot v}) \biggl[
{1 \over 2}v^iv^j \left( {m \over r}(3-8\eta)-2v^2(1-5\eta) \right)
-n^{(i}v^{j)}{m \over r}\dot r (7+4\eta)
\nonumber \\
&&
- n^in^j{m \over r}\biggl( {3 \over 4}(1-2\eta)\dot r^2
+ {1 \over 3}(26-3\eta){m \over r} - {1 \over 4}(7-2\eta)v^2 \biggr)
\biggr] \,, \\
P^{3/2}Q_{SO}^{ij} = && {2 \over r^2} \biggl\{ n^in^j \left[ ({\bf
\hat n
\times v}){\bf \cdot}(12{\bf S} +6 {\delta m \over m}{\bf
\Delta})\right]
- n^{(i}\left[{\bf v \times}(9{\bf S}+5{\delta m \over m}{\bf
\Delta})\right]
^{j)} \nonumber \\ &&
+ \dot r n^{(i}\left[{\bf \hat n \times}
(12{\bf S}+6{\delta m \over m}{\bf \Delta})
\right]^{j)}
- v^{(i}\left[{\bf \hat n \times}(2{\bf S}+2{\delta m \over m}{\bf
\Delta})
\right]^{j)} \nonumber \\ &&
+ 3 \dot r ({\bf \hat N \cdot \hat n}) \left[ ({\bf S}+{\delta m \over
m}
{\bf \Delta}){\bf \times \hat N} \right]^{(i} n^{j)}
- 2  ({\bf \hat N \cdot v}) \left[ ({\bf S}+{\delta m \over m}
{\bf \Delta}){\bf \times \hat N} \right]^{(i} n^{j)} \nonumber \\ &&
- 2  ({\bf \hat N \cdot \hat n}) \left[ ({\bf S}+{\delta m \over m}
{\bf \Delta}){\bf \times \hat N} \right]^{(i} v^{j)} \biggr\} \,, \\
P^{3/2}Q_{Tail}^{ij}= && 4m \int_0^\infty \biggl\{ {m \over r^3} \left [
(3v^2+{m \over r}-5 \dot r^2 )n^in^j+18 \dot r n^{(i}v^{j)}-4v^iv^j
\right ] \biggr\}_{u-u^\prime} \nonumber \\
&& \times \left [ \ln \left ( {u^\prime \over 2s}
\right ) + {11 \over 12} \right ] du^\prime \,.
\end{eqnarray}
For all but the tail contributions, equivalent results follow from the
less rigorous methods of Epstein and Wagoner (1975) and Wagoner and
Will (1976).

\begin{figure}[t]
   \vspace{27pc}
   \caption{Gravitational waveform plotted against orbital phase for
an equal-mass system, showing the quadrupole and tail contributions.  Plotted
is $(R/2\mu)h_+$, viewed by an observer in the orbital plane.  Plots
begin at an orbital separation of $18 m$ and terminate around $6 m$
(from Wiseman 1993).}
\end{figure}

\section{Energy and Angular Momentum Loss}

The radiative energy and angular momentum loss in terms of STF radiative
multipoles are given, to the order of accuracy we require, by

\begin{equation}
{dE \over dt} = - {1 \over 5} \left\{ \stackrel{(3)}{I_{ij}}
 \stackrel{(3)}{I_{ij}} + {5 \over 189} \stackrel{(4)}{I_{ijk}}
 \stackrel{(4)}{I_{ijk}} + {16 \over 9} \stackrel{(3)}{J_{ij}}
\stackrel{(3)}{J_{ij}} \right\}\,,
\end{equation}
\begin{equation}
{dJ^i \over dt} = - {2 \over 5} \epsilon_{ijk} \left\{
\stackrel{(2)}{I_{jl}}
 \stackrel{(3)}{I_{kl}} + {5 \over 126} \stackrel{(3)}{I_{jlm}}
 \stackrel{(4)}{I_{klm}} + {16 \over 9} \stackrel{(2)}{J_{jl}}
\stackrel{(3)}{J_{kl}} \right\} \,.
\end{equation}
Taking the appropriate time derivatives of the STF moments
and using the post-Newtonian
equations of motion to the necessary order, one finds, for the energy
flux,
\begin{equation}
{dE \over dt} = {\dot E}_N + {\dot E}_{PN} + {\dot E}_{SO} + {\dot
E}_{Tail}+ O(\epsilon^2){\dot E}_N  \,,
\end{equation}
where
\begin{eqnarray}
{\dot E}_N = && - {8 \over 15} {m^2 \mu^2 \over r^4} \left\{ 12v^2 -
11 \dot
r^2 \right\} \,, \\
{\dot E}_{PN} = && - {8 \over 15} {m^2 \mu^2 \over r^4} \biggl\{ {1
\over 28}
\biggl[ (785-852\eta)v^4 - 2(1487-1392\eta)v^2 \dot r^2 +
3(687-620\eta) \dot
r^4 \nonumber \\ && - 160(17-\eta) {m \over r}v^2
+ 8(367-15\eta){m \over r} \dot r^2 +
16(1-4\eta)({m \over r})^2 \biggr] \biggr\} \,, \\
{\dot E}_{SO} = && - {8 \over 15} {m^2 \mu^2 \over r^4} \biggl\{ {{\bf
\hat n
\times v} \over mr} {\bf \cdot} \biggl[ {\bf S} (78 \dot r^2 -80v^2
-8
{m \over r})
+ {{\delta m} \over m} {\bf \Delta} (51 \dot r^2 - 43v^2 + 4{m \over r})
\biggr ]
\biggr\} , \\
{\dot E}_{Tail}= && {4m \over 5} \stackrel{(4)}{I^{ij}}(u)
\int_0^\infty \stackrel{(4)}{I^{ij}}(u-u^\prime) \ln u^\prime du^\prime
\,,
\end{eqnarray}
and for the angular momentum flux
\begin{equation}
{d{\bf J} \over dt} = {\dot {\bf J}}_N + {\dot {\bf J}}_{PN}
+ {\dot {\bf J}}_{SO} + {\dot {\bf J}}_{Tail} + O(\epsilon^2){\dot {\bf
J}}_N \,,
\end{equation}
where
\begin{eqnarray}
{\dot {\bf J}}_N = && - {8 \over 5} {m \mu^2 \over r^3}
{\bf {\tilde L}}_N
\left\{ 2v^2 - 3 \dot r^2 + 2 {m \over r} \right\} \,, \\
{\dot {\bf J}}_{PN} = && - {8 \over 5} {m \mu^2 \over r^3}
{\bf {\tilde L}}_N
\biggl\{ {1 \over 84}
\biggl[ (307-548\eta)v^4 - 6(74-277\eta)v^2 \dot r^2 + 15(19-72\eta)
\dot
r^4 \nonumber \\ && - 4(58+95\eta) {m \over r}v^2
+ 2(372+197\eta){m \over r} \dot r^2 -
2(745-2\eta)({m \over r})^2 \biggr] \biggr\} \,, \\
{\bf \dot J}_{SO} = && -{4 \over 5} {\mu^2 \over r^3} \biggl\{
{1 \over r^2}
{\bf {\tilde L}}_N {\bf {\tilde L}}_N {\bf \cdot} \left[ (65 \dot r^2
-37v^2
-{163 \over 3}{m \over r}){\bf S} + (35 \dot r^2 -19v^2 -{71 \over 3}
{m \over r}){\delta m \over m}{\bf \Delta} \right] \nonumber \\ &&
+ {m \over r}{\bf \hat n \times} \left[ ({\bf \hat n \times S})(15
\dot r^2
-{41 \over 3}v^2 +{4 \over 3}{m \over r}) + {\delta m \over m}
({\bf \hat n \times \Delta}) (9\dot r^2 - 8v^2 -{2 \over 3}{m \over
r}) \right]
\nonumber \\ &&
+ \dot r {\bf v \times} \left[ ({\bf \hat n \times S})(18 {m \over r}
+ 44v^2 -55 \dot r^2) + {\delta m \over m}
({\bf \hat n \times \Delta}) ({25 \over 3}{m \over r} +20v^2-25\dot
r^2) \right]
\nonumber \\ &&
+ {\bf v \times} \left[ ({\bf v \times S})(36 \dot r^2
-{71 \over 3}v^2 -{50 \over 3}{m \over r}) + {\delta m \over m}
({\bf v \times \Delta}) (18\dot r^2 - {35 \over 3}v^2-9{m \over r})
\right]
\nonumber \\ &&
- \dot r {m \over r}{\bf \hat n \times} \left[ 4({\bf v \times S}) +
{5 \over 3}{\delta m \over m}({\bf v \times \Delta}) \right]
+ {2 \over 3}{m \over r}(\dot r^2 - v^2){\delta m \over m}{\bf \Delta}
\biggr\} \,, \\
{\dot J}_{Tail}^i = && {8m \over 5} \epsilon^{ijk}
\stackrel{(3)}{I^{jl}}(u) \int_0^\infty \stackrel{(4)}{I^{kl}}(u-u^\prime)
\ln u^\prime du^\prime  \,,
\end{eqnarray}
where ${\bf {\tilde L}}_N \equiv {\bf x \times v}$.
Because the tail contribution depends in principle on the past history
of the orbit (retarded times $u^\prime$ prior to $u$), it can only
be evaluated explicitly for specific orbits, such as quasi-circular orbits,
or head-on collisions.

\begin{figure}[t]
   \vspace{25pc}
   \caption{Modulation of gravitational waveforms by spin-induced
orbital precession, plotted against time to coalescence.
System consists of a non-spinning neutron star
and a maximally spinning black hole of $10 M_\odot$.  Spin and orbital
angular momentum vectors are initially misaligned by $\alpha=11.3^o$.
Initial orbital inclination relative to $\bf J$ is $i$.  Angle
$\gamma$ represents orientation of interferometer relative to $\bf J$,
with both $\bf J$ and the detector plane assumed parallel to the plane
of the sky.  Curves show envelope of waveform for various detector
orientations (absolute height of curves has been shifted for ease of
presentation).  Gravitational-wave frequency runs from 10 Hz on the
right to 300 Hz on the left.  Numbers running from
25 to 250 indicate the number of gravity-wave cycles per modulation
(from Kidder 1994).}
\end{figure}

\section{Circular Orbits}

Here we specialize to quasi-circular orbits, that is, orbits that are
circular on an orbital timescale, but that inspiral on a
radiation-reaction timescale.  These will be most relevant for the
inspiral phase of systems that have evolved from a previous bound
orbit since, in general, such orbits circularize long before they
reach the final plunge state (see Lincoln and Will (1990) for discussion).
By truncating
the radiation-reaction terms in the equations of motion, we can
characterize circular orbits by $\ddot r=\dot r =0$.  Using the fact
that $\ddot r={\bf \hat n \cdot a}+r \omega^2$, where
$\omega=d\phi/dt$ is the angular velocity, we have, from Eqs. (6) --
(8), for a circular
orbit,
\begin{equation}
v^2=(r \omega)^2={m \over r} \left\{ 1-(3-\eta){m \over r} - {1 \over
{mr^2}} {\bf \tilde L}_N  \cdot (5 {\bf S}+3 {{\delta m} \over m} {\bf
\Delta}) \right\} \label{omegacirc} \,,
\end{equation}
where we keep only post-Newtonian and spin-orbit terms.
A circular orbit does not exist in
general when spin-spin interactions are included, except in an
orbit-averaged sense (see Kidder {\it et al.} (1993b) for discussion).
The energy and angular momentum are
given by
\begin{eqnarray}
\tilde E &&= -{1 \over 2}{m \over r} + {1 \over 8}(7-\eta)({m \over
r})^2-{1 \over {2r^3}} {\bf \tilde L}_N \cdot (3{\bf S}+{{\delta m}
\over m} {\bf \Delta}) \,, \label{Ecirc} \\
{\bf {\tilde J}} &&={\bf S}/\mu +{\bf {\tilde L}}_N \left\{ 1+ {1 \over 2}
(7-\eta)
{m \over r} \right\} \nonumber \\
&&+ {1 \over r} \left\{ {\bf {\hat n}} \times \left
[  {\bf {\hat n}} \times (3{\bf S} + {{\delta m} \over m} {\bf
\Delta}) \right ] - {1 \over 2} {\bf {\hat \lambda}} \times \left
[  {\bf {\hat \lambda}} \times ({\bf S} + {{\delta m} \over m} {\bf
\Delta}) \right ] \right\} \,, \label{Jcirc}
\end{eqnarray}
where ${\tilde E} \equiv E/\mu$, ${\bf {\tilde J}}\equiv {\bf J}/\mu$
and ${\bf {\hat \lambda}}= {\bf v}/v$.
For a circular orbit the energy and angular momentum fluxes are given
by
\begin{eqnarray}
\dot E= &&-{32 \over 5} \eta^2 ({m \over r})^5 \biggl\{ 1- {1 \over
336}(2927 +420\eta){m \over r}+4\pi({m \over r})^{3/2}
\nonumber \\
&&-{1 \over 12} ({m \over r})^{3/2}{\bf {\hat L}}_N \cdot (148{\bf S}+75
{{\delta m}
\over m} {\bf \Delta})/m^2
 \biggr\} \,, \label{dEcirc} \\
{\bf {\dot J}}= && -{32 \over 5} \eta^2 ({m \over r})^4 r^2 \omega
\biggl\{ {\bf {\hat L}}_N \biggl [ 1- {1 \over 336} (1919+756\eta){m
\over r} + 4\pi ({m \over r})^{3/2} \nonumber \\
&&- {1 \over 12} ({m \over r})^{3/2}
{\bf {\hat L}}_N \cdot (88 {\bf S}+39{{\delta m} \over m} {\bf
\Delta})/m^2 \biggr ] \nonumber \\
&&+ {1 \over 24} ({m \over r })^{3/2} \left [ {\bf {\hat \lambda}}{\bf
{\hat \lambda}} \cdot (37{\bf S}+24 {{\delta m} \over m} {\bf
\Delta})/m^2 +
{\bf {\hat n}}{\bf
{\hat n}} \cdot (121{\bf S}+60 {{\delta m} \over m} {\bf \Delta})/m^2
\right ] \biggr\} \,, \label{dJcirc}
\end{eqnarray}
where
${\bf {\tilde L}}_N =r^2\omega {\bf {\hat L}}_N$.  The term
$4\pi(m/r)^{3/2}$ is the tail effect.  It is assumed
that the individual spinning objects are sufficiently
axisymmetric that they emit negligible gravitational radiation on
account of their internal motions, and, in particular that their spins
are unaffected by radiation damping, to the order considered.
In the test body ($\eta=0$) limit, Eq. (\ref{dEcirc}) agrees with results of
calculations of perturbations of Kerr and Schwarzschild spacetimes (Cutler,
Finn,
{\it et al.} 1993, Poisson
1993a,b, Tagoshi and Nakamura 1994) .
In fact, Tagoshi and Nakamura (1994) have carried out test-body
calculations to order
$(m/r)^4$, including $(m/r)^4 \ln (m/r) $ terms.
Differentiating Eq. (\ref{Ecirc}),
and substituting Eqs. (\ref{dEcirc}) and
(\ref{omegacirc}) we obtain, for
the evolution of the orbital radius
\begin{eqnarray}
\dot r=&&-{64 \over 5}\eta ({m \over r})^3 \biggl\{ 1-{1 \over
336}(1751+588\eta){m \over r} +4\pi ({m \over r})^{3/2} \nonumber \\
&& -{7 \over
12}({m \over r})^{3/2} {\bf {\hat L}}_N \cdot (34{\bf S}+15{{\delta m} \over m}
{\bf \Delta})/m^2  \biggr\} \,.
\end{eqnarray}
It is useful to note that, for circular orbits, $\dot E= \omega {\bf
{\hat L}}_N \cdot {\bf {\dot J}}$.
Equations (\ref{omegacirc}), (\ref{Ecirc}) and (\ref{dEcirc}) can
also be used to express $\dot \omega$ in terms of
$\omega$.  When written in terms of the gravitational-wave frequency,
$f=\omega/\pi$, one finds
\begin{eqnarray}
{{\dot f} \over f^2} =&& {96\pi \over 5} \eta m^{5/3} (\pi f)^{5/3}
\biggl\{ 1- {1 \over 336} (743+924\eta)(\pi mf)^{2/3} +4\pi(\pi mf)
\nonumber \\
&& - {1 \over 12}(\pi mf) {\bf {\hat L}}_N \cdot (188 {\bf S}+75
{{\delta m} \over m} {\bf \Delta})/m^2 \biggr\} \,. \label{fdot}
\end{eqnarray}
It is conventional to define the ``chirp mass'', ${\cal M} \equiv
\eta^{3/5} m = (m_1m_2)^{3/5}m^{-1/5}$; this parameter governs the
leading, quadrupole-order evolution of $f$, so that ${\dot f}/f^2
\approx (96\pi/5)(\pi {\cal M} f)^{5/3} $.

\section{Results}

Here we present some representative results obtained using these
post-Newtonian methods.  Figure 1 shows inspiralling, quasi-circular
orbits in the late stage leading up to the plunge for an equal-mass
system, and for a system with a 10:1 mass ratio.  Figure 2 shows the
relative size of post-Newtonian effects on the waveform for a system
of spinless bodies with a 10:1 mass ratio, evolving from $r=15m$ to
$r=4m$.
In Figure 3, the relative size of the tail term is shown for an
equal-mass system, evolving from $r=18m$ to $r=6m$.  In Figure 4, the
effect of spin-induced precession of the orbital plane on the waveform
is shown for a system containing a non-spinning neutron star and a
maximally spinning ($S_2=m_2^2$), $10 M_\odot$ black hole, with the
spin misaligned with the orbital angular momentum.   In all cases, the
observer is in the plane perpendicular to the total angular momentum.

As was first pointed out by Cutler and Finn (see Cutler {\it et al.}
(1993) for discussion), the use of theoretical templates having the
frequency evolution built in via Eq. (\ref{fdot}), matched against the
hundreds to thousands of waves detected in the sensitive bandwidth,
will permit accurate determinations of the system parameters, such as
the chirp mass, the reduced mass and the spins.  Cutler and Flanagan
(1994) and Finn and Chernoff (1993) have shown that the proposed
phase-sensitive template matching can be a powerful tool for
determining the important parameters of the coalescing system.  For
example, Table 5 (from Cutler and Flanagan (1994))
shows the accuracies that could be achieved for the
chirp mass $\cal M$ and the reduced mass $\mu$, when the first
post-Newtonian corrections (terms proportional to $(\pi mf)^{2/3}$) in
Eq. (\ref{fdot}) are taken into account, ignoring spin effects.

\vspace{1 pc}

Table 5.  Estimating Parameters of Coalescing Binaries

\begin{tabular}{llll} \\ \hline
$m_1$&$m_2$&$\Delta {\cal M}/{\cal M}$&$\Delta \mu/\mu$ \\ \hline
1.4&1.4&0.0040 \%&0.41 \% \\
1.4&10&0.020 \%&0.54 \% \\
10&10&0.16 \%&1.9 \% \\ \hline
\end{tabular}

\vspace{1pc}

When spin effects are included, the accuracy of determination of $\mu$
degrades substantially, because of the strong correlation between the
$f$-dependence of the two corresponding terms in Eq. (\ref{fdot}).  To
date, such estimates have not taken into account spin-induced modulation of the
waveforms, which may provide another handle on the effects of spins.  On
the other hand, spin effects in double-neutron star systems are not
likely to be important (or measurable), either in the frequency
evolution or in waveform modulations, because their spins are
typically small compared to $m^2$, and thus small compared to the
orbital angular momentum.  Modulations are likely to be more important
for systems containing rapidly spinning black holes (see Apostolatos {\it
et
al.} 1994, Kidder 1994 for further discussion).

\section{Acknowledgments}

This research was supported in part by the National Science Foundation
under Grant No. PHY 92-22902, and the National Aeronautics and Space
Administration under Grants No. NAGW 2902 \& 3874.  I am grateful to
Lawrence Kidder for especially useful discussions and for careful
checking of formulas.  Alan Wiseman, Bala Iyer and Luc Blanchet also
provided useful comments.

This paper was an invited talk at the 8th Nishinomiya-Yukawa
Symposium, held in Nishinomiya, Japan, October 28-29, 1993.  I am
grateful to the organizers and the city of Nishinomiya for their
support.

\section{References}

\re
Abramovici, A., {\it et al.}, 1992, {\it Science} {\bf 256}, 325.
\re
Andersson, N., 1992, {\it Proc. R. Soc. London} {\bf 439}, 47.
\re
Anninos, P., Hobill, D., Seidel, E., Smarr, L., and Suen, W.-M., 1993,
{\it Phys. Rev. Lett.} {\bf 71}, 2581.
\re
Apostolatos, T. A., Cutler, C., Sussman, G. J. and Thorne, K. S.,
1994, {\it Phys. Rev. D}, submitted.
\re
Bildsten, L. and Cutler, C., 1992, {\it Astrophys. J.} {\bf 400}, 175.
\re
Blanchet, L. and Damour, T., 1986, {\it Philos. Trans. R. Soc. London}
{\bf A320}, 379.
\re
Blanchet, L. and Damour, T., 1988, {\it Phys. Rev. D} {\bf 37}, 1410.
\re
Blanchet, L. and Damour, T., 1989, {\it Ann. Inst. H. Poincar\`e} {\bf 50},
377.
\re
Blanchet, L. and Damour, T., 1992, {\it Phys. Rev. D} {\bf 46}, 4304.
\re
Blanchet, L., Damour, T., Iyer, B. R., Will, C. M., and Wiseman, A.
G., 1994, work in progress.
\re
Blanchet, L. and Sch\"afer, G., 1989, {\it Mon. Not. R. Astron. Soc.}
{\bf 239}, 845.
\re
Centrella, J. and McMillan, S., 1993, {\it Astrophys. J.} {\bf 416},
719.
\re
Chernoff, D. F. and Finn, L. S. 1993, {\it Astrophys. J. Lett.} {\bf
411}, L5.
\re
Christodoulou, D., 1991, {\it Phys. Rev. Lett.} {\bf 67}, 1486.
\re
Clark, J. P. A. and Eardley, D. M., 1977, {\it Astrophys. J.} {\bf
215}, 311.
\re
Clark, J. P. A., van den Heuvel, E. P. J., and Sutantyo, W., 1979,
{\it Astron. Astrophys.} {\bf 72}, 120.
\re
Cutler, C., Apostolatos, T. A., Bildsten, L., Finn, L. S., Flanagan,
\'E. E., Kennefick, D., Markovi\'c, D. M., Ori, A., Poisson, E.,
Sussman, G. J., and Thorne, K. S., 1993, {\it Phys. Rev. Lett.} {\bf
70}, 2984.
\re
Cutler, C., Finn, L. S., Poisson, E., and Sussman, G. J., 1993, {\it
Phys. Rev. D} {\bf 47}, 1511.
\re
Cutler, C. and Flanagan, \'E. E., 1994, {\it Phys. Rev. D}, in press.
\re
Damour, T., 1987, in {\it 300 Years of Gravitation}, edited by S. W.
Hawking and W. Israel (Cambridge University Press, Cambridge), p.
128.
\re
Damour, T. and Iyer, B. R., 1991, {\it Ann. Inst. H. Poincar\`e} {\bf 54}, 115.
\re
Detweiler, S. L. and Blackburn, K., 1992, {\it Phys. Rev. D} {\bf 46},
2318.
\re
Dyson, F. J., 1963, in {\it Interstellar Communication}, edited by A. G.
W. Cameron (Benjamin, New York), p. 115.
\re
Epstein, R. and Wagoner, R. V., 1975, {\it Astrophys. J.} {\bf 197},
717.
\re
Finn, L. S., 1992, {\it Phys. Rev. D} {\bf 46}, 5236.
\re
Finn, L. S. and Chernoff, D. F., 1993, {\it Phys. Rev. D} {\bf 47}, 2198.
\re
Iyer, B. R. and Will, C. M., 1993, {\it Phys. Rev. Lett.} {\bf 70},
116.
\re
Iyer, B. R. and Will, C. M., 1994, in preparation.
\re
Jotania, K., Wagh, S. J., and Dhurandhar, S. V., 1992, {\it Phys. Rev.
D} {\bf 46}, 2507.
\re
Junker, W. and Sch\"afer, G., 1992, {\it Mon. Not. R. Astron. Soc.} {
\bf 254}, 146.
\re
Kidder, L. E., 1994, in preparation.
\re
Kidder, L. E., Will, C. M., and Wiseman, A. G., 1992, {\it Class.
Quantum Grav. Lett.} {\bf 9}, L125.
\re
Kidder, L. E., Will, C. M., and Wiseman, A. G., 1993a, {\it Phys. Rev.
D} {\bf 47}, 3281.
\re
Kidder, L. E., Will, C. M., and Wiseman, A. G., 1993b, {\it Phys. Rev.
D} {\bf 47}, R4183.
\re
Krolak, A., Lobo, J. A., and Meers, B. J., 1991, {\it Phys. Rev. D}
{\bf 43}, 2470.
\re
Krolak, A., Lobo, J. A., and Meers, B. J., 1993, {\it Phys. Rev. D}
{\bf 47}, 2184.
\re
Krolak, A. and Schutz, B. F., 1987, {\it Gen. Relativ. Gravit.} {\bf
19}, 1163.
\re
\re
Lattimer, J. M. and Schramm, D. N., 1974, {\it Astrophys. J.
Lett.} {\bf 192}, L145.
\re
Lattimer, J. M. and Schramm, D. N., 1976, {\it Astrophys. J.} {\bf 210}, 549.
\re
Lincoln, C. W. and Will, C. M., 1990, {\it Phys. Rev. D} {\bf 42},
1123.
\re
Markovi\'c, D. M., 1993, {\it Phys. Rev. D} {\bf 48}, 4738.
\re
Narayan, R., Piran, T., Shemi, A., 1991, {\it Astrophys. J. Lett.} {\bf 379},
L17.
\re
Phinney, E. S., 1991, {\it Astrophys. J. Lett.} {\bf 380},
L17.
\re
Poisson, E., 1993a, {\it Phys. Rev. D} {\bf 47}, 1497.
\re
Poisson, E., 1993b, {\it Phys. Rev. D} {\bf 48}, 1860.
\re
Rasio, F. A. and Shapiro, S. L., 1992, {\it Astrophys. J.} {\bf 401},
226.
\re
Schutz, B. F., 1986, {\it Nature} {\bf 323}, 310.
\re
Schutz, B. F., 1991, in {\it The Detection of Gravitational Waves},
edited by D. G. Blair (Cambridge University Press, Cambridge), p.
406.
\re
Shibata, M., Nakamura, T., and Oohara, K.-I., 1992, {\it Prog. Theor.
Phys.} {\bf 88}, 1079.
\re
Tagoshi, H. and Nakamura, T., 1994, {\it Phys. Rev. D}, in press.
\re
Thorne, K. S., 1987, in {\it 300 Years of Gravitation}, edited by S. W.
Hawking and W. Israel (Cambridge University Press, Cambridge), p.
128.
\re
Thorne, K. S., 1992, {\it Phys. Rev. D} {\bf 45}, 520.
\re
Wagoner, R. V. and Will, C. M., 1976, {\it Astrophys. J.} {\bf 210},
764; {\it ibid.} {\bf 215}, 984 (1977).
\re
Wiseman, A. G.,  1992, {\it Phys. Rev. D} {\bf 46}, 1517.
\re
Wiseman, A. G.,  1993, {\it Phys. Rev. D} {\bf 48}, 4757.
\re
Wiseman, A. G. and Will, C. M.,  1991, {\it Phys. Rev. D} {\bf 44}, R2945.

\end{document}